\documentclass[prd]{revtex4}

\usepackage{bm}

\def\ba{{\bm a}}
\def\bb{{\bm b}}
\def\bh{{\bm h}}
\def\bk{{\bm k}}
\def\bl{{\bm l}}
\def\bn{{\bm n}}
\def\bv{{\bm v}}
\def\bw{{\bm w}}

\def\bx{{\bm x}}
\def\bxp{{\bm x'}}
\def\bxpp{{\bm x''}}
\def\by{{\bm y}}
\def\bD{{\bm D}}
\def\bN{{\bm N}}
\def\bR{{\bm R}}
\def\bS{{\bm S}}
\def\bW{{\bm W}}

\def\al{\alpha}
\def\l{\lambda}

\def\Omv{\Omega(x_{A}, x_{B})}
\def\Om{\Omega}
\def\bom{{\bm \omega}}
\def\bna{{\bm \nabla}}
\def\bla{{\bm \lambda}}

\def\lb{\label}

\def\be{\begin{equation}}
\def\ee{\end{equation}}
\def\bea{\begin{eqnarray}}
\def\eea{\end{eqnarray}}

\begin{document}

\title{A universal tool for determining the time delay and the frequency shift of light: \\
Synge's world function} 
 
\author{Pierre Teyssandier}
\email{Pierre.Teyssandier@obspm.fr}
\affiliation{D\'epartement Syst\`emes de R\'ef\'erence Temps et Espace,
CNRS/UMR 8630, \\
Observatoire de Paris, 61 avenue de l'Observatoire, F-75014 Paris, France}

\author{Christophe Le Poncin-Lafitte}
\email{leponcin@danof.obspm.fr}
\affiliation{D\'epartement Syst\`emes de R\'ef\'erence Temps et Espace,
CNRS/UMR 8630, \\
Observatoire de Paris, 61 avenue de l'Observatoire, F-75014 Paris, France}

\author{Bernard Linet}
\email{linet@celfi.phys.univ-tours.fr}
\affiliation{Laboratoire de Math\'ematiques et Physique Th\'eorique, 
CNRS/UMR 6083, Universit\'e Fran\c{c}ois Rabelais, F-37200 Tours, France}

\date{\today}

\begin{abstract}

In almost all of the studies devoted to the time delay and the frequency shift of light, the calculations are based on the integration of the null geodesic equations. However, the above-mentioned effects can be calculated without integrating the geodesic equations if one is able to determine the bifunction $\Omega(x_A, x_B)$ giving half the squared geodesic distance between two points $x_A$ and $x_B$ (this bifunction may be called Synge's world function). In this lecture, $\Omega(x_A, x_B)$ is determined up to the order $1/c^3$ within the framework of the PPN formalism. The case of a stationary gravitational field generated by an isolated, slowly rotating axisymmetric body is studied in detail. The calculation of the time delay and the frequency shift is carried out up to the order $1/c^4$. Explicit formulae are obtained for the contributions of the mass, of the quadrupole moment and of the internal angular momentum when the only post-Newtonian parameters different from zero are $\beta$ and $\gamma$. It is shown that the frequency shift induced by the mass quadrupole moment of the Earth at the order $1/c^3$ will amount to $10^{-16}$ in spatial experiments like the ESA's Atomic Clock Ensemble in Space mission. Other contributions are briefly discussed.

\end{abstract}


\maketitle

\section{Introduction}

A lot of fundamental tests of gravitational theories rest on highly precise measurements of the travel time 
and/or the frequency shift of electromagnetic signals propagating through the gravitational field of the 
Solar System. In practically all of the previous studies, the explicit expressions of such travel times 
and frequency shifts as predicted by various metric theories of gravity are derived from an integration of 
the null geodesic differential equations. 
This method works quite well within the first post-Minkowskian approximation, as it is shown by the results obtained, e.g.,
in \cite{kli1, kop1, kop2, kli3, kop3}. Of course, it works also within the post-Newtonian approximation, especially in the case of a static, spherically symmetric space-time treated up to order $1/c^3$ \cite{ashby,bla1}. However, the solution of the geodesic equations requires heavy calculations when one has to take into account the presence of mass multipoles in the field or the tidal effects due to the planetary motions, and the calculations become quite complicated in the post-post-Minkowskian approximation \cite{richter2}, especially in the dynamical case \cite{brugmann}. 

The aim of this lecture is to present a quite different procedure recently developed by two of us. Based on Synge's world function \cite{syn}, this procedure avoids the integration of the null geodesic equations and is particularly convenient for determining the light rays which connect an emitter and a receiver having specified spatial locations at a finite distance. Thus, we are able to extend the previous calculations of the time delay and of the frequency shift up to the order $1/c^4$. As a consequence, it is now possible to predict the time/frequency transfers in the vicinity of the Earth at a level of accuracy which amounts to $10^{-18}$ in fractional frequency. This level of accuracy is expected to be reached in the foreseeable future with optical atomic clocks \cite{hol}. 

The plan of the lecture is as follows. First, in Sect. II, the definition of the time transfer functions are given and the invariant expression of the frequency shift is recalled. It is shown that explicit expressions of the frequency shift can be derived when the time transfer functions are known. 
In Sect. III, the relevant properties of Synge's world function are recalled. In Sect. IV, the general expressions of the 
world function and of the time transfer functions are obtained within the Nordtvedt-Will parametrized post-Newtonian (PPN) formalism. In Sect. V, the 
case of a stationary field generated by an isolated, slowly rotating axisymmetric body is analyzed in detail. It is shown that the contributions of the mass and spin multipoles can be obtained by straightforward derivations of a single function. Retaining only the terms due to the 
mass $M$, to the quadrupole moment $J_2$ and to the intrinsic angular momentum $\bS$ of the rotating body, explicit expansions of the 
world function and of the time transfer function are derived up to the order $1/c^3$ and $1/c^4$, respectively. The same formalism yields the vectors tangent to the light ray at the emitter and at the receiver up to the order $1/c^3$. In Sect. VI, the frequency shift is developed up to the order $1/c^4$ on the assumption that $\beta$ and $\gamma$ are 
the only non vanishing post-Newtonian parameters. Explicit expressions are obtained for the contributions of $J_2$ 
and $\bS$. Numerical estimates are given for the ESA's Atomic Clock Ensemble in Space (ACES) mission \cite{sal,spa}. Concluding remarks 
are given in Sect. VII.

Equivalent results formulated with slightly different notations may be found in \cite{linet1} and an extension of the method to the general post-Minkowskian approximation is given in \cite{leponcin1}.

{\em Notations}. -- In this work, $G$ is the Newtonian gravitational constant and $c$ is the 
speed of light in a vacuum. The Lorentzian metric of space-time is denoted 
by $g$. The signature adopted for $g$ is $ (+ - - -)$. We suppose that the 
space-time is covered by one global coordinate system 
$(x^{\mu})=(x^0,\bx )$, where $x^0=ct$, $t$ being a time coordinate, and
$\bx =(x^i)$, the $x^i$ being quasi Cartesian coordinates. We choose coordinates $x^i$ so that 
the curves of equations $x^{i} = $ const are timelike. This choice means that $g_{00} > 0$ 
everywhere. We employ the vector 
notation $\ba$ in order to denote either  $\{a^1, a^2, a^3\} = \{a^i\}$ or
$\{a_1, a_2, a_3\} = \{a_i\}$. 
Considering two such quantities $\ba$ and $\bb$ 
with for instance $\ba = \{a^i\}$, we use $\ba \cdot \bb$ to denote 
$a^i b^i$ if $\bb= \{b^i\}$ or $a^i b_i$ if $\bb = \{b_i\}$ 
(the Einstein convention on the repeated indices is used). The quantity 
$\mid \! \ba \! \mid$ stands for the ordinary Euclidean norm of $\ba$.

\section{Time transfer functions, time delay and frequency shift}

We consider here electromagnetic signals propagating through a vacuum between an emitter $A$ and a receiver $B$. We suppose that these signals may be assimilated to light rays travelling along null geodesics of the metric (geometic optics approximation). We call $x_A$ the point of emission by $A$ 
and $x_B$ the point of reception by $B$. We put $x_A = (ct_A, \bx_A)$ and $x_B = (ct_B, \bx_B)$. We assume that there do not exist two distinct null geodesics  starting from $x_A$ and intersecting the world line of $B$. These assumptions are clearly satisfied in all experiments currently envisaged in the Solar System. 

1. {\em Time transfer functions and time delay}.-- The quantity $t_B - t_A$ is the (coordinate) travel time of the signal. Upon the above mentioned assumptions, $t_B - t_A$ may be considered either as a function of the instant of emission $t_A$ and of  $\bx_A$, $\bx_B$, or as a function of the instant of reception $t_B$ and of $\bx_A$ and $\bx_B$. So, we can in general define two distinct (coordinate) time transfer functions, ${\cal T}_e$ and ${\cal T}_r$ by putting  :
\be \lb{ttf}
t_B - t_A = {\mathcal T}_e(t_A, \bx_A, \bx_B) = {\mathcal T}_r(t_B, \bx_A, \bx_B) \, .
\ee
We call ${\cal T}_e$ the emission time transfer function and ${\cal T}_r$ the reception time transfer function. As we shall see below, 
the main problem will consist in 
determining explicitly these functions when the metric is given. Of course, it is in principle sufficient to determine one of these functions.

We shall put
\be \lb{R}
R_{AB} = \, \mid\!\bx_B - \bx_A\!\mid \, .
\ee
throughout this work. The time delay is then defined as $t_B - t_A - R_{AB}/c$. It is well known that this quantity is $>0$ in Schwarzschild 
space-time, which explains its designation \cite{sha}.

2. {\em Frequency shift}.-- Denote by $u_{A}^{\alpha}$ and $u_{B}^{\alpha}$ the unit 4-velocity vectors of the emitter at $x_A$ and of the receiver at $x_B$, respectively. Let $\Gamma_{AB}$ be the null geodesic path connecting $x_A$ and $x_B$, described by parametric equations $x^{\alpha} = x^{\alpha}(\zeta)$, $\zeta$ being an affine parameter. Denote by $l^{\mu}$ the vector tangent to $\Gamma_{AB}$ defined as 
\be \lb{tg}
l^{\mu} = \frac{dx^{\mu}}{d\zeta} \, .
\ee 
Let $\nu_A$ be the frequency of the signal emitted at $x_A$ as measured by a clock comoving with $A$ and $\nu_B$  be the frequency of the same signal received at $x_B$ as measured by a clock comoving with $B$. The ratio $\nu_A/\nu_B$ is given by the well-known formula \cite{syn}
\be \lb{56}
\frac{\nu_A}{\nu_B} = \frac{u_{A}^{\mu}(l_{\mu})_A}{u_{B}^{\mu}(l_{\mu})_B} \, .
\ee 
Since it is assumed that the emission and reception points are connected by a single null geodesic, it is clear that $(l_{\mu})_A$ and 
$(l_{\mu})_B$ may be considered either as functions of the instant of emission $t_A$ and of  $\bx_A$, $\bx_B$, or as functions of the instant of reception $t_B$ and of $\bx_A$ and $\bx_B$. Therefore, we may write  
\be \lb{FT}
\frac{\nu_A}{\nu_B} = {\mathcal N}_e(u_A, u_B; t_A, \bx_A, \bx_B) = {\mathcal N}_r(u_A, u_B; t_B, \bx_A, \bx_B)\, .
\ee

Denote by $\bv_A=(d\bx/dt)_A$ and $\bv_B=(d\bx/dt)_B$ the
coordinate velocities of the observers at $x_A$ and $x_B$, respectively: 
\be \lb{vel}
\bv_A = \left(\frac{d\bx}{dt}\right)_A \, , \quad \bv_B = \left(\frac{d\bx}{dt}\right)_B \, .
\ee
It is easy to see that the formula (\ref{56}) may be written as
\be \lb{57}
\frac{\nu_A}{\nu_B}=\frac{u_{A}^{0}}{u_{B}^{0}}\, \frac{(l_0)_A}{(l_0)_B}\, \frac{q_A}{q_B} \, ,\quad
q_A=1+\frac{1}{c}\widehat{\bl}_A\cdot \bv_A \, , \quad 
q_B=1+\frac{1}{c}\widehat{\bl}_B\cdot \bv_B \, ,
\ee
where $\widehat{\bl}_A$ and $\widehat{\bl}_B$ are the quantities defined as
\be \lb{hl}
\widehat{\bl}_A = \left\{\left(\frac{l_i}{l_0}\right)_A\right\}\, , \qquad 
\widehat{\bl}_B = \left\{\left(\frac{l_i}{l_0}\right)_B\right\} \, .
\ee

It is immediately deduced from Eq. (\ref{57}) that an explicit expression of ${\mathcal N}_e$ (resp. ${\mathcal N}_r$) can be derived when the time transfer function ${\mathcal T}_e$ (resp. ${\mathcal T}_r$) is known. Indeed, one has the following theorem \cite{leponcin1}.

\medskip
{\bf Theorem}.  {\em Consider a signal emitted at point} $x_A = (ct_{A}, \bx_{A})$ 
{\em and received at point} $x_B = (ct_{B}, \bx_{B})$. {\em Denote by} $l^{\mu}$ 
{\em the vector} $dx^{\mu}/d\zeta$ {\em tangent to the null geodesic at point} $x(\zeta)$, $\zeta$ {\em being any affine parameter, and put}
\be \lb{hat}
\widehat{l}_i = \left(\frac{l_{i}}{l_{0}}\right) \, .
\ee

{\em Then, one has relations as follow at} $x_A$ {\em and at} $x_B$   
\bea 
& &\left(\widehat{l}_i\right)_A = 
c \, \frac{\partial  {\cal T}_{e}}{\partial x^{i}_{A}}
\left[1 + \frac{\partial  {\cal T}_{e}}{\partial t_{A}}
\right]^{-1} \, = \, c \, \frac{\partial {\cal T}_{r}}{\partial x^{i}_{A}} \, , \lb{2d2} \\
& & \nonumber \\
& &\left(\widehat{l}_i\right)_B = 
-c \, \frac{\partial {\cal T}_{e}}{\partial x^{i}_{B}} \, = \,
- c \, \frac{\partial  {\cal T}_{r}}{\partial x^{i}_{B}}
\left[1 - \frac{\partial  {\cal T}_{r}} {\partial t_B}\right]^{-1}\,  , \lb{2d1} \\
& & \nonumber \\
& &\frac{(l_{0})_A}{(l_{0})_B} =  1 + 
\frac{\partial  {\cal T}_{e}}{\partial t_{A}} \, 
= \, \left[1 -
\frac{\partial  {\cal T}_{r}}{\partial t_{B}}\right]^{-1}   \, , \lb{2d3}
\eea
\noindent
{\em where} ${\cal T}_{e}$ {\em and} ${\cal T}_{r}$ {\em are taken at} 
$(t_A, \bx_A, \bx_B)$ {\em and} $(t_B, \bx_A, \bx_B)$, {\em respectively}.

This theorem may be straightforwardly deduced from a fundamental property of the world function that we 
introduce in the following section.

{\em Case of a stationary space-time}. In a stationary space-time, we can choose coordinates
$(x^{\mu})$ such that the metric does not depend on $x^0$. Then, the travel time of the signal only depends on $\bx_A, \bx_B$. 
This means that Eq. (\ref{ttf}) reduces to a single relation of the form 
\be \lb{4}
t_B-t_A = {\cal T}(\bx_A,\bx_B) \, .
\ee
It immediately follows from Eqs. (\ref{2d2}) and (\ref{2d1}) that
\bea 
(\widehat{l}_i)_A &=&
c\frac{\partial}{\partial x_{A}^{i}}{\cal T}(\bx_A,\bx_B) \, , \lb{6} \\
(\widehat{l}_i)_B &=&
-c\frac{\partial}{\partial x_{B}^{i}}{\cal T}(\bx_A,\bx_B) \, , \lb{5} \\
\frac{(l_0)_A}{(l_0)_B} &=& 1\, . \lb{AB1}
\eea

As a consequence, the formula (\ref{57}) reduces now to
\be \lb{nu0}
\frac{\nu_A}{\nu_B} = \frac{u_{A}^{0}}{u_{B}^{0}}\, 
\frac{1 + \bv_A \cdot \bna_{\bx_A} {\cal T}}{1 - \bv_B \cdot \bna_{\bx_B} {\cal T}} \, ,
\ee
where $\bna_{\bx} f$ denotes the usual gradient operator acting on $f(\bx)$.

It is worthy of note that $(1, \{(\widehat{l}_i)_A\})$ and $(1, \{(\widehat{l}_i)_B\})$ constitute a set of covariant components of the 
vector tangent to the light ray at $\bx_A$ and $\bx_B$, respectively. This tangent vector corresponds to the affine parameter chosen 
so that $(l_0)_A = (l_0)_B =1$.

\section{The world function and its post-Newtonian limit}

\subsection{Definition and fundamental properties}

For a moment, consider $x_A$ and $x_B$ as arbitrary points. We assume that there exists one and only one geodesic path, 
say $\Gamma_{AB}$, which links these two points. This assumption means that point $x_B$ belongs to the 
normal convex neighbourhood \cite{poisson} of point $x_A$ (and conversely that 
$x_A$ belongs to the normal convex neighbourhood of point $x_B$). The world function is 
the two-point function $\Omega (x_A,x_B)$ defined by
\be \lb{Om}
\Omega (x_A,x_B) = \frac{1}{2} \epsilon_{AB} [s_{AB}]^2 \, ,
\ee
where $s_{AB}$ is the geodesic distance between $x_A$ and $x_B$, namely
\be \lb{geo}
s_{AB} = \int_{\Gamma_{AB}} \sqrt{g_{\mu \nu}dx^{\mu} dx^{\nu}}
\ee
and $\epsilon_{AB} = 1, 0, -1$ according as $\Gamma_{AB}$ is a timelike, a null or a spacelike geodesic. 
An elementary calculation shows that $\Omega (x_A,x_B)$ may be written in any case as \cite{syn} 
\be \lb{1} 
\Om (x_A,x_B)=\frac{1}{2}\int_{0}^{1}g_{\mu \nu}(x^{\alpha}(\lambda ))
\frac{dx^{\mu}}{d\lambda}\frac{dx^{\nu}}{d\lambda}d\lambda \, , 
\ee
where the integral is taken along $\Gamma_{AB}$, $\lambda$ denoting the unique affine parameter 
along $\Gamma_{AB}$ which fulfills the boundary conditions $\lambda_A=0$ and 
$\lambda_B=1$.
 
It follows from Eqs. (\ref{Om}) or (\ref{1}) that the world function 
$\Om (x_A,x_B)$ is unchanged if we perform any admissible coordinate 
transformation. 

The utility of the world function for our purpose comes from the following properties \cite{syn,leponcin1}. 

{\em i}) The vectors $(dx^{\alpha}/d\l)_A$ and $(dx^{\alpha}/d\l)_B$ tangent to the 
geodesic $\Gamma_{AB}$ respectively at $x_{A}$ and $x_{B}$ are given by
\be \lb{3}
\left( g_{\alpha \beta}\frac{dx^{\beta}}{d\lambda}\right)_A =
-\frac{\partial \Omega}{\partial x_{A}^{\alpha}} \, , \quad
\left( g_{\alpha \beta}\frac{dx^{\beta}}{d\lambda}\right)_B =
\frac{\partial \Omega}{\partial x_{B}^{\alpha}} \, . 
\ee

As a consequence, if $\Omega (x_A,x_B)$ is explicitly known, the determination of these vectors 
does not require the integration of the differential equations of the geodesic. 

{\em ii}) Two points $x_A$ and $x_B$ are linked by a null geodesic if and 
only if the condition 
\be \lb{2}
\Om (x_A,x_B)=0 
\ee
is fulfilled. Thus, $\Om (x_A,x)=0$ is the equation of the null cone ${\cal C}(x_{A})$ at $x_A$. 

Consequently, if the bifunction $\Om (x_A,x_B)$ is explicitly known, it is in principle possible to determine the 
emission time transfer function ${\mathcal T}_e$ by solving the equation 
\be \lb{tt}
\Om (ct_A, \bx_A, ct_B, \bx_B) = 0
\ee
for $t_B$. It must be pointed out, however, that solving Eq. (\ref{tt}) for $t_B$ 
yields two distinct solutions $t_B^{+}$ and $t_B^{-}$ since the timelike 
curve $ x^{i} = x_B^{i}$ cuts the light cone ${\cal C}(x_{A})$ at two points $x_B^{+}$ and $x_B^{-}$, 
$x_B^{+}$ being in the future of $x_B^{-}$. Since we regard $x_A$ as the 
point of emission of the signal and $x_B$ as the point of reception, we shall exclusively focus our attention on the 
determination of $t_{B}^{+} - t_{A}$ (clearly, the determination of $t_{B}^{-} - t_{A}$ comes within 
the same methodology). For the sake of brevity, we shall henceforth write $t_B$ instead of $t_{B}^{+}$. 

Of course, solving Eq. (\ref{tt}) for $t_A$ yields the reception time transfer function ${\mathcal T}_r$.

Generally, extracting the time transfer fonctions from Eq. (\ref{tt}), next using Eqs. (\ref{2d2}) or (\ref{2d1}) will be more 
straightforward than deriving the vectors tangent at $x_A$ and $x_B$ 
from (\ref{3}), next imposing the constraint (\ref{2}). 

To finish, note that the theorem stated in Sect. II is easily deduced from the identities
\be 
\Omega(ct_A, \bx_A, ct_A + c{\mathcal T}_e (t_A, \bx_A, \bx_B), \bx_B) \equiv 0  \nonumber 
\ee
and
\be
\Omega(ct_B - c{\mathcal T}_r (t_B, \bx_A, \bx_B), \bx_A, ct_B, \bx_B) \equiv 0 \, . \nonumber 
\ee

\subsection{General expression of the world function in the post-Newtonian limit}

We assume that the metric may be written as 
\be \lb{8}
g_{\mu \nu}=\eta_{\mu \nu} + h_{\mu \nu}
\ee
throughout space-time, with $\eta_{\mu \nu} = $diag$ (1, -1, -1, -1)$. Let $\Gamma_{AB}^{(0)}$ be the 
straight line defined by the parametric equations 
$x^{\al} = x_{(0)}^{\al}(\l)$, with
\be \lb{9}
x_{(0)}^{\alpha}(\lambda) = (x_{B}^{\alpha}-x_{A}^{\alpha}) \lambda +
x_{A}^{\alpha} \, , \quad 0 \leq \l \leq 1 \, .
\ee
With this definition, the parametric equations of the geodesic $\Gamma_{AB}$ 
connecting $x_A$ and $x_B$ may be written in the form
\be \lb{9a}
x^{\al}(\l) = x_{(0)}^{\al}(\l) + X^{\al}(\l) \, , \quad 0 \leq \l \leq 1  \, ,
\ee
where the quantities $X^{\al}(\l)$ satisfy the boundary conditions
\be \lb{9b}
X^{\al}(0) = 0 \, , \quad  X^{\al}(1) = 0 \, .
\ee
Inserting Eq. (\ref{8}) and $dx^{\mu}(\l)/d\l = x_{B}^{\mu}-x_{A}^{\mu} + dX^{\mu}(\l)/d\l$ 
in Eq. (\ref{1}), then developing and noting that
\be \nonumber  
\int_{0}^{1}\eta_{\mu \nu}(x_B^{\mu} - x_A^{\mu}) \frac{dX^{\nu}}{d\l} d\l = 0
\ee
by virtue of Eq. (\ref{9b}), we find the rigorous formula
\bea \lb{9c}
\Omv & = & \Om^{(0)}(x_A,x_B) + \frac{1}{2} (x_B^{\mu} - x_A^{\mu})(x_B^{\nu} - x_A^{\nu})
\int_{0}^{1} h_{\mu \nu}(x^{\alpha}(\lambda)) d \l \nonumber \\
&  & \mbox{} + \frac{1}{2} \int_{0}^{1} \left[ g_{\mu\nu}(x^{\al}(\l)) \frac{dX^{\mu}}{d\l}
\frac{dX^{\nu}}{d\l} + 
2 (x_B^{\mu} - x_A^{\mu})h_{\mu\nu}(x^{\al}(\l)) \frac{dX^{\nu}}{d\l} \right] d\l \, , 
\eea
where the integrals are taken over $\Gamma_{AB}$ and $ \Om^{(0)}(x_A,x_B)$ is the 
world function in Minkowski space-time
\be \lb{7}
\Om^{(0)}(x_A,x_B)=\frac{1}{2}\eta_{\mu \nu}(x_{B}^{\mu}-x_{A}^{\mu})
(x_{B}^{\nu}-x_{A}^{\nu}) \, .
\ee

Henceforth, we shall only consider weak gravitational fields generated by 
self-gravitating extended bodies within the slow-motion, post-Newtonian 
approximation. So, we assume that the potentials $h_{\mu\nu}$ may be 
expanded as follows 
\bea \lb{1n}
& & h_{00}=\frac{1}{c^2}h_{00}^{(2)}+\frac{1}{c^4}h_{00}^{(4)}+O(6) \nonumber \, , \\
& & h_{0i}=\frac{1}{c^3}h_{0i}^{(3)}+O(5) \, , \\ 
& & h_{ij}=\frac{1}{c^2}h_{ij}^{(2)}+O(4) \, . \nonumber
\eea
From these expansions and from the Euler-Lagrange equations satisfied by any geodesic curve, namely
\be \lb{2n}
\frac{d}{d\lambda}\left( g_{\alpha \beta}\frac{dx^{\beta}}{d\lambda}
\right) =\frac{1}{2}\partial_{\alpha}h_{\mu \nu}
\frac{dx^{\mu}}{d\lambda}\frac{dx^{\nu}}{d\lambda} \, ,
\ee 
it results that $X^{\mu}(\l) = O(2)$ and that $dx^{\mu}/d\l = x_{B}^{\mu}-x_{A}^{\mu} + O(2)$. As a 
consequence, $h_{\mu\nu}(x^{\al}(\l)) = h_{\mu\nu}(x_{(0)}^{\al}(\l)) + O(4)$ and the 
third and fourth terms in the r.h.s. of Eq. (\ref{9c}) are of order $1/c^4$. These features result in an 
expression for $\Om (x_A,x_B)$ as follows
\be \lb{10}
\Om (x_A,x_B)=\Om^{(0)}(x_A,x_B)+\Om^{(PN)}(x_A,x_B) + O(4) \, ,
\ee
where
\bea \lb{4n}
& & \Omega^{(PN)}(x_A,x_B) = \frac{1}{2c^2}(x_{B}^{0}-x_{A}^{0})^2
\int_{0}^{1}h_{00}^{(2)}(x_{(0)}^{\alpha}(\lambda ))d\lambda \nonumber \\
& & \qquad \qquad \qquad \quad +\frac{1}{2c^2}(x_{B}^{i}-x_{A}^{i})
(x_{B}^{j}-x_{A}^{j})\int_{0}^{1}h_{ij}^{(2)}(x_{(0)}^{\alpha}(\lambda ))d\lambda \nonumber \\
& & \qquad \qquad \qquad \quad +\frac{1}{c^3}(x_{B}^{0}-x_{A}^{0})
(x_{B}^{i}-x_{A}^{i})\int_{0}^{1}h_{0i}^{(3)}(x_{(0)}^{\alpha}(\lambda ))
d\lambda \, , 
\eea
the integral being now taken over the line $\Gamma_{AB}^{(0)}$ defined by Eq. (\ref{9}).

The formulae (\ref{10}) and (\ref{4n}) yield the general expression of the
world function up to the order $1/c^3$ within the framework of the 1 PN
approximation. We shall see in the next subsection that this approximation
is sufficient to determine the time transfer functions up to the order $1/c^4$. 
It is worthy of note that the method used above would as well lead to the expression of the 
world function in the linearized weak-field limit previously found by Synge \cite{syn}.

\subsection{Time transfer functions at the order $1/c^4$}

Suppose that $x_B$ is the point of reception of a signal emitted at $x_A$. Taking Eq. (\ref{10}) into 
account, Eq. (\ref{2}) may be written in the form
\[ 
\Om^{(0)}(x_A,x_B)+\Om^{(PN)}(x_A,x_B)=O(4) \, ,
\]
which implies the relation
\be \lb{12}
t_{B} - t_{A} = \frac{1}{c} R_{AB} - \frac{\Omega^{(PN)}(ct_A, \bx_A , ct_B, \bx_B)}{c R_{AB}}+ O(4) \, .
\ee
Using iteratively this relation, we find for the emission time transfer function
\be \lb{5n}
{\cal T}_{e}(t_{A}, \bx_{A}, \bx_{B}) = \frac{1}{c} R_{AB}
-\frac{\Omega^{(PN)}(ct_A ,\bx_A , ct_A + R_{AB}, \bx_B )}{cR_{AB}} + O(5) \, .
\ee
and for the reception time transfer function
\be \lb{5n1}
{\cal T}_{r}(t_{B}, \bx_{A}, \bx_{B}) = \frac{1}{c} R_{AB}
-\frac{\Omega^{(PN)}(ct_B - R_{AB} ,\bx_A , ct_B, \bx_B )}{cR_{AB}} + O(5) \, .
\ee
These last formulae show that the time transfer functions can be 
explicitly calculated up to
the order $1/c^4$ when $\Omega^{(PN)}(x_A,x_B)$ is known. This fundamental
result will be exploited in the following sections. 

It is worthy of note that a comparison of Eqs. (\ref{5n}) and (\ref{5n1}) immediately gives the following relations :
\be \lb{Tr}
{\cal T}_{r}(t_{B}, \bx_{A}, \bx_{B}) = {\cal T}_{e}\left(t_{B}- \frac{R_{AB}}{c}, \bx_{A}, \bx_{B}\right) + O(5)
\ee 
and conversely
\be \lb{Te}
{\cal T}_{e}(t_{A}, \bx_{A}, \bx_{B}) = {\cal T}_{r}\left(t_{A} + \frac{R_{AB}}{c} , \bx_{A}, \bx_{B}\right) + O(5) \, .
\ee

The quantity $\Omega^{(PN)}(ct_A ,\bx_A , ct_A + R_{AB}, \bx_B )$ in (\ref{5n}) may be 
written in an integral form by using Eq. (\ref{4n}), in which $R_{AB}$ and $R_{AB}\l + ct_A$ are 
substituted for $x_B^{0} - x_A^{0}$ and for $x_{(0)}^{0}(\l)$, respectively. As a consequence
\be \lb{14}
{\cal T}_{e}(t_{A}, \bx_{A}, \bx_{B}) = \frac{1}{c}R_{AB}\left\{ 1-\frac{1}{2c^2}\int_{0}^{1}
\left[ h_{00}^{(2)}(z_{+}^{\al}(\l)) + h_{ij}^{(2)}(z_{+}^{\al}(\l))N^i N^j 
+ \frac{2}{c}h_{0i}^{(3)}(z_{+}^{\al}(\l))N^i \right] d\lambda \right\} +O(5) \, ,
\ee
the integral being taken over curve $\Gamma_{AB}^{(0)+}$ defined by the parametric equations 
$x^{\al} = z_{+}^{\al}(\l)$, where 
\be \lb{13}
z_{+}^0(\lambda )=R_{AB}\lambda + ct_A \, , \quad
z_{+}^i(\lambda )=R_{AB}N^i\lambda +x_{A}^{i} \, , \quad 0\leq \lambda \leq 1 \, ,
\ee
with
\be \lb{N}
R_{AB}=\mid \! \bR_{AB}\! \mid \, , \qquad N^i = \frac{x_{B}^{i} - x_{A}^{i}}{R_{AB}} \, .
\ee
We note that $\Gamma_{AB}^{(0)+}$ is a null geodesic path of Minkowski 
metric from $x_A$, having the above-defined quantities $N^{i}$ as direction cosines. 

A similar reasoning leads to an expression as follows for ${\cal T}_{r}$
\be \lb{14r}
{\cal T}_{r}(t_{B}, \bx_{A}, \bx_{B}) = \frac{1}{c}R_{AB}\left\{ 1-\frac{1}{2c^2}\int_{0}^{1}
\left[ h_{00}^{(2)}(z_{-}^{\al}(\l)) + h_{ij}^{(2)}(z_{-}^{\al}(\l))N^i N^j 
+ \frac{2}{c}h_{0i}^{(3)}(z_{-}^{\al}(\l))N^i \right] d\lambda \right\} +O(5) \, ,
\ee
the integral being now taken over curve $\Gamma_{AB}^{(0)-}$ defined by the parametric equations 
$x^{\al} = z_{-}^{\al}(\l)$, where 
\be \lb{13r}
z_{-}^0(\lambda ) = - R_{AB}\lambda + ct_B \, , \quad
z_{-}^i(\lambda ) = - R_{AB}N^i\lambda +x_{B}^{i} \, , \quad 0\leq \lambda \leq 1 \, .
\ee
Curve $\Gamma_{AB}^{(0)-}$ is a null geodesic path of Minkowski metric arriving at $x_B$ and 
having $N^i$ as direction cosines.

\section{World function and time transfer functions within the Nordtvedt-Will PPN 
formalism}

\subsection{Metric in the 1 PN approximation}

In this section, we use the Nordvedt-Will post-Newtonian formalism involving 
ten parameters $\beta$, $\gamma$, $\xi$, $\alpha_1$, $\ldots$, $\zeta_4$
\cite{will}. We introduce slightly modified notations in order to be closed of the 
formalism recently proposed by Klioner and Soffel \cite{kli4}
as an extension of the post-Newtonian framework elaborated by Damour, Soffel and 
Xu \cite{dam} for general relativity. In particular, we denote by 
$\bv_r$ the velocity of the center of mass O relative to the universe
rest frame \footnote{This velocity is noted $\bw$ in Ref. \cite{will}.}.

Although our method is not confined to any particular assumption on the
matter, we suppose here that each source of the field is described by the
energy-momentum tensor of a perfect fluid
\[
T^{\mu \nu}=\rho c^2\left[ 1+\frac{1}{c^2}\left( \Pi +\frac{p}{\rho}\right)
\right] u^{\mu}u^{\nu}-pg^{\mu \nu} \, ,
\]
where $\rho$ is the rest mass density, $\Pi$ is the specific energy density
(ratio of internal energy density to rest mass density), $p$ is the
pressure and $u^{\mu}$ is the unit 4-velocity of the fluid. In this section
and in the following one, $\bv$ is the coordinate velocity $d\bx /dt$ of
an element of the fluid. We introduce the conserved mass density $\rho^*$
given by
\be \lb{M4}
\rho^*=\rho\sqrt{-g}u^0=\rho \left[ 1+\frac{1}{c^2}\left( \frac{1}{2}v^2
+3\gamma U \right) +O(4) \right] \, ,
\ee
where $g = \det (g_{\mu\nu})$ and $U$ is the Newtonian-like potential
\be \lb{M5}
U(x^0,\bx )=G\int \frac{\rho^*(x^0,\bxp )}{\mid \! \bx -\bxp \! \mid}
d^3\bxp\,  . 
\ee

In order to obtain a more simple form than the usual one for the potentials
$h_{0i}$, we suppose that the chosen $(x^{\mu})$ are related to a standard
post-Newtonian gauge $(\overline{x}^{\mu})$ by the transformation
\be \lb{M2}
x^0=\overline{x}^0 + \frac{1}{c^3}\left[ (1+ 2\xi +\alpha_2-\zeta_1 )
\partial_t\chi
-2\alpha_2\bv_r\cdot \bna \chi \right] , \quad x^i=\overline{x}^i \, ,
\ee
where $\chi$ is the superpotential defined by
\be \lb{M3}
\chi (x^0,\bx )=\frac{1}{2} G \int \rho^*(x^0,\bxp )\mid \! \bx -\bxp \! \mid 
d^3\bxp \, .
\ee
Moreover, we define $\widehat{\rho}$ by 
\bea \lb{M6}
\widehat{\rho} & = & \rho^* \left[ 1+\frac{1}{2}(2\gamma +1-2\xi +\alpha_3 +\zeta_1)
\frac{v^2}{c^2} +(1-2\beta +\xi +\zeta_2)\frac{U}{c^2}+(1+\zeta_3)
\frac{\Pi}{c^2}+(3\gamma -2\xi +3\zeta_4 )\frac{p}{\rho^*c^2} \right.
\nonumber \\
&   &  \left. \mbox{} \quad \quad \quad \quad \quad -\frac{1}{2}(\alpha_1 -\alpha_3)\frac{v^{2}_{r}}{c^2}
-\frac{1}{2}(\alpha_1-2\alpha_3) \frac{\bv_r \cdot \bv}{c^2} + O(4) \right]  \, .
\eea 
Then, the post-Newtonian potentials read
\bea 
& & h_{00}=-\frac{2}{c^2}w+\frac{2\beta}{c^4}w^2+\frac{2\xi}{c^4}\phi_W
+\frac{1}{c^4}(\zeta_1-2\xi )\phi_v-\frac{2\alpha_2}{c^4}
v_{r}^{i}v_{r}^{j}\partial_{ij}\chi +O(6) , \lb{M7} \\
& & \bh \equiv \{h_{0i}\} =\frac{2}{c^3}\left[ \left( \gamma +1 +\frac{1}{4}
\alpha_1 \right) \bw +\frac{1}{4}\alpha_1 w \, \bv_r \right] +O(5) , \lb{M8} \\
& & h_{ij}=-\frac{2\gamma}{c^2}w\delta_{ij} +O(4) \, , \lb{M9}
\eea
where
\bea 
& & w(x^0,\bx ) = G \int \frac{\widehat{\rho}(x^0,\bxp )}{\mid \! \bx -\bxp \! \mid }d^3\bxp
+\frac{1}{c^2}\left[ (1+ 2\xi +\alpha_2 -\zeta_1 )\partial_{tt}\chi
-2\alpha_2\bv_r \cdot \bna (\partial_t\chi ) \right] , \lb{M10} \\
& & \phi_W(x^0,\bx )=G^2\int \frac{\rho^*(x^0,\bxp )\rho^*(x^0,\bxpp )(\bx -\bxp )}
{\mid \! \bx -\bxp \! \mid^3}\cdot \left( \frac{\bxp -\bxpp}
{\mid \! \bx -\bxpp \! \mid}-\frac{\bx -\bxpp}{\mid \! \bxp -\bxpp \! \mid}
\right) d^3\bxp d^3\bxpp , \lb{M11} \\
& & \phi_v(x^0,\bx )=G\int \frac{\rho^*(x^0,\bxp )[\bv (x^0,\bxp )\cdot 
(\bx -\bxp )]^2}{\mid \! \bx -\bxp \! \mid^3}d^3\bxp , \lb{M12} \\
& & \bw (x^0,\bx )=G\int \frac{\rho^*(x^0,\bxp )\bv (x^0,\bxp )}
{\mid \! \bx -\bxp \! \mid}d^3\bxp \, . \lb{M13} 
\eea

\subsection{Determination of the world function and of the time transfer functions}

For the post-Newtonian metric given by Eqs. (\ref{M7})-(\ref{M13}),
it follows from Eq. (\ref{4n}) that $\Omega (x_A,x_B)$ may be written 
up to the order $1/c^3$ in the form given by Eq. (\ref{10}) with
\be \lb{25}
\Om^{(PN)}(x_A,x_B)=\Om^{(PN)}_{w}(x_A,x_B)+\Om^{(PN)}_{\bw}(x_A,x_B)
+\Om^{(PN)}_{\bv_r}(x_A,x_B) \, ,
\ee
where
\bea 
& & \Om^{(PN)}_{w}(x_A,x_B)=-\frac{1}{c^2} \left[ (x^{0}_{B}-x^{0}_{A})^2
+\gamma R_{AB}^{2}\right] \int_{0}^{1}w(x_{(0)}^{\alpha}(\lambda ))
d\lambda \, ,\lb{25a} \\
& & \Om^{(PN)}_{\bw}(x_A,x_B)=\frac{2}{c^3}\left( \gamma +1 + \frac{1}{4}\alpha_1 \right) 
(x^{0}_{B}-x^{0}_{A})\bR_{AB}\cdot 
\int_{0}^{1}\bw (x^{\alpha}_{(0)}(\lambda ))d\lambda \, , \lb{25b} \\
& & \Om^{(PN)}_{\bv_r}(x_A,x_B)=\frac{1}{2c^3}\alpha_1 (x^{0}_{B}-x^{0}_{A})
(\bR_{AB}\cdot \bv_r )\int_{0}^{1}w(x^{\alpha}_{(0)}(\lambda ))d\lambda \, ,
\lb{25c}
\eea
the integrals being calculated along the curve defined by Eq. (\ref{9}).

The emission time transfer function is easily obtained by using Eqs. (\ref{5n}) or 
(\ref{14}). We get
\bea \lb{6n}
{\cal T}_{e}(t_{A}, \bx_{A}, \bx_{B}) & = & \frac{1}{c}R_{AB} + \frac{1}{c^3}(\gamma +1)R_{AB}
\int_{0}^{1}w(z_{+}^{\al}(\l)) d\lambda  \nonumber \\
& - & \frac{2}{c^4}\bR_{AB}\cdot \left[ (\gamma +1+\frac{1}{4}\alpha_1 )
\int_{0}^{1}\bw (z_{+}^{\al}(\l)) d\lambda +\frac{1}{4} \alpha_{1}
\bv_r \int_{0}^{1}w(z_{+}^{\al}(\l)) d\lambda \right] +O(5) \, ,
\eea
the integral being evaluated along the curve $\Gamma_{AB}^{(0)+}$ defined by Eq. (\ref{13}).

The reception time transfer function is given by 
\bea \lb{6nr}
{\cal T}_{r}(t_{B}, \bx_{A}, \bx_{B}) & = & \frac{1}{c}R_{AB} + \frac{1}{c^3}(\gamma +1)R_{AB}
\int_{0}^{1}w(z_{-}^{\al}(\l)) d\lambda  \nonumber \\
& - & \frac{2}{c^4}\bR_{AB}\cdot \left[ (\gamma +1+\frac{1}{4}\alpha_1 )
\int_{0}^{1}\bw (z_{-}^{\al}(\l)) d\lambda +\frac{1}{4} \alpha_{1}
\bv_r \int_{0}^{1}w(z_{-}^{\al}(\l)) d\lambda \right] +O(5) \, ,
\eea
the integral being evaluated along the curve $\Gamma_{AB}^{(0)-}$ defined by Eq. (\ref{13r}).

Let us emphasize that, since $w=U+O(2)$, $w$ may be replaced by the
Newtonian-like potential $U$ in Eqs. (\ref{25a})-(\ref{6n}).

\subsection{Case of an stationary source}

In what follows, we suppose that the gravitational field is generated by a
single stationary source. Then, $\partial_t\chi=0$ and the potentials $w$ and 
$\bw$ do not depend on time. In this case, the integration involved in 
Eqs. (\ref{25a})-(\ref{25c}) can be  performed by a method due to Buchdahl 
\cite{buc1}. Introducing the auxiliary variables $\by_A=\bx_A-\bxp$ and
$\by_B=\bx_B-\bxp$, and replacing in Eq. (\ref{9}) the parameter $\lambda$ by
$u=\lambda -1/2$, a straightforward calculation yields
\bea 
& & \int_{0}^{1}w(\bx_{(0)}(\lambda ))d\lambda =G \int \widehat{\rho}(\bxp )
F(\bxp ,\bx_A,\bx_B)d^3\bxp , \lb{28} \\
& & \int_{0}^{1}\bw (\bx_{(0)}(\lambda ))d\lambda =
G \int \rho^*(\bxp ) \bv (\bxp ) F(\bxp ,\bx_A,\bx_B)d^3\bxp \, , \lb{29}
\eea  
where the kernel function $F(\bxp ,\bx_A,\bx_B)$ has the expression
\be \lb{KF}
F(\bxp  ,\bx_A,\bx_B)=\int_{-1/2}^{1/2}\frac{du}
{\mid \! (\by_B-\by_A)u+\frac{1}{2}(\by_B+\by_A)\! \mid} \, .
\ee
Noting that $\by_B-\by_A=\bR_{AB}$, which implies that
$\mid \! \by_B-\by_A\! \mid =R_{AB}$, we find
\be \lb{19}
F(\bx ,\bx_A,\bx_B)=\frac{1}{R_{AB}}\ln \left( \frac{\mid \! \bx -\bx_A\! \mid
+\mid \! \bx -\bx_B\! \mid+R_{AB}}{\mid \! \bx -\bx_A\! \mid +
\mid \! \bx -\bx_B\! \mid -R_{AB}} \right) \, .
\ee 

Inserting Eqs. (\ref{28}), (\ref{29}) and (\ref{19}) in Eqs. (\ref{25a})-(\ref{25c}) and in 
Eq. (\ref{6n}) will enable one to obtain quite elegant expressions for $\Om^{(PN)}(x_A,x_B) $ 
and for ${\cal T}(\bx_{A}, \bx_{B})$, respectively.

\section{Isolated, slowly rotating axisymmetric body}

Henceforth, we suppose that the light is propagating in the gravitational
field of an isolated, slowly rotating axisymmetric body. The gravitational field
is assumed to be stationary. The main purpose of this section is to
determine the influence of the mass and spin multipole moments of the rotating
body on the coordinate time transfer and on the direction of light rays. 
From these results, it will be possible to obtain a relativistic modelling of 
the one-way time transfers and frequency shifts up to the order $1/c^4$ in a 
geocentric non rotating frame.

Since we treat the case of a body located very far from the other bodies
of the universe, the global coordinate system $(x^{\mu})$ used until now can
be considered as a local (i.e. geocentric) one. So, in agreement with the 
UAI/UGG Resolution B1 (2000) \cite{uai}, we shall henceforth denote by $W$ and 
$\bW$ the quantities $w$ and $\bw$ respectively defined by
Eqs. (\ref{M10}) and (\ref{M13}) and we shall denote by
$G_{\mu \nu}$ the components of the metric. However, we shall continue here 
with using lower case letters for the geocentric coordinates in order to 
avoid too heavy notations.

The center of mass O of the rotating body being taken as the origin of the 
quasi Cartesian coordinates $(\bx )$, we choose the axis of symmetry as the 
$x^3$-axis. We assume that the body is rotating about O$x^3$ with a 
constant angular velocity $\bom$, so that 
\be \lb{34a}
\bv (\bx )=\bom \times \bx \, .
\ee
In what follows, we put $r=\mid \! \bx \! \mid$, $r_A=\mid \! \bx_A \! \mid$
and $r_B=\mid \! \bx_A \! \mid$. We call $\theta$ the angle between $\bx$ and 
O$x^3$. We consider only the case where all points of the segment joining 
$\bx_A$ and $\bx_B$ are outside the body. We denote by $r_e$ the radius of the 
smallest sphere centered on O and containing the body (for celestial bodies, $r_e$ is the
equatorial radius). In this section, we assume the convergence of the multipole 
expansions formally derived below at any point outside the body, even if $r<r_e$.  

\subsection{Multipole expansions of $W$ and $\bW$}

According to Eqs. (\ref{M10}), (\ref{M13}) and (\ref{34a}), the gravitational potentials
$W$ and $\bW$ obey the equations
\be \lb{M14}
\bna^2 W=-4\pi G\widehat{\rho} \, , \quad 
\bna^2 \bW =-4\pi G\rho^* \bom \times \bx \, .
\ee
It follows from Eq. (\ref{M14}) that the potential 
$W$ is a harmonic function outside the rotating body. As a consequence, $W$ may be expanded 
in a multipole series of the form
\be \lb{31}
W(\bx )=\frac{GM}{r}\left[ 1-\sum_{n=2}^{\infty}J_n
\left( \frac{r_e}{r}\right)^n P_n(\cos \theta )\right] \, .
\ee
In this expansion, $P_n$ is the Legendre polynomial of degree $n$ and the 
quantities $M$, $J_2, \ldots,J_{n}, \ldots$ correspond to the
generalized Blanchet-Damour mass multipole moments in general relativity 
\cite{bla2}.

For the sake of simplicity, put
\[
z = x^3 \, .
\]
Taking into account the identity
\[
\frac{\partial^n}{\partial z^n}\left( \frac{1}{r}\right)=
\frac{(-1)^nn!}{r^{1+n}}P_n(z/r) \, ,\quad z=x^3 \, ,
\]  
it may be seen that
\be \lb{32}
W(\bx ) = GM\left[ \frac{1}{r}-\sum_{n=2}^{\infty}\frac{(-1)^n}{n!}J_n
r_{e}^{n}\frac{\partial^n}{\partial z^n}\left( \frac{1}{r}\right) \right]\, .
\ee
Substituting for $W$ from Eq. (\ref{32}) into Eq. (\ref{M14}) yields an expansion for 
$\widehat{\rho}$ as follows
\be \lb{33}
\widehat{\rho}(\bx )=M\left[ \delta^{(3)}(\bx )-\sum_{n=2}^{\infty}\frac{(-1)^n}{n!}
J_nr_{e}^{n}\frac{\partial^n}{\partial z^n} \delta^{(3)}(\bx ) \right] \, ,
\ee
where $\delta^{(3)}(\bx )$ is the Dirac distribution supported by the origin O. This expansion 
of $\widehat{\rho}$ in a multipole series will be exploited in the next subsection.

Now, substituting Eq. (\ref{34a}) into Eq. (\ref{M13}) yields for the vector potential $\bW$
\be \lb{34b}
\bW (\bx ) = G \int \frac{\rho^* (\bxp )\bom \times \bxp}
{\mid \! \bx -\bxp \! \mid}d^3\bxp \, .
\ee
It is possible to show that this vector may be written as
\be \lb{34}
\bW=-\frac{1}{2}\bom \times \bna \cal V \, ,
\ee
where ${\cal V}$ is an axisymmetric function satisfying the Laplace
equation $\bna^2 {\cal V}=0$ outside the body. 
Consequently, we can expand $\cal V$ in a series of the form
\be \lb{35}
{\mathcal V}(\bx )=\frac{GI}{r}\left[ 1-\sum_{n=1}^{\infty}K_n
\left( \frac{r_e}{r}\right)^n P_n(\cos \theta )\right] \, ,
\ee 
where $I$ and each $K_n$ are constants. Substituting for ${\mathcal V}$ from Eq. (\ref{35}) into 
Eq. (\ref{34}), and then using the identity
\[
(n+1)P_n(z/r)+(z/r)P'_n(z/r)=P'_{n+1}(z/r) \, ,
\] 
we find an expansion for $\bW$ as follows
\be \lb{36}
\bW (\bx )=\frac{GI\bom \times \bx}{2r^3}\left[ 1 -\sum_{n=1}^{\infty}
K_n\left( \frac{r_e}{r}\right)^nP'_{n+1}(\cos \theta ) \right] \, ,
\ee
which coincides with a result previously obtained by one of us \cite{tey}. This coincidence 
shows that $I$ is the moment of inertia of the body about the $z$-axis. Thus, the 
quantity $\bS=I\bom$ is the intrinsic angular momentum of the rotating body. The coefficients 
$K_n$ are completely determined by the density distribution $\rho^*$ and by the shape of 
the body \cite{tey, adler}. The quantities $I, K_1, K_2, ...K_n, ...$ correspond to the 
Blanchet-Damour spin multipoles in the special case of a stationary axisymmetric 
gravitational field.

Equation (\ref{36}) may also be written as 
\be \lb{37}
\bW (\bx )=-\frac{1}{2}G\bS \times \bna \left[ \frac{1}{r}
-\sum_{n=1}^{\infty}\frac{(-1)^n}{n!}K_nr_{e}^{n}
\frac{\partial^n}{\partial z^n}\left( \frac{1}{r}\right) \right] \, .
\ee
Consequently, the density of mass current can be expanded in the
multipole series
\be \lb{38}
\rho^*(\bx )(\bom \times \bx )=-\frac{1}{2}\bS \times \bna \left[
\delta^{(3)}(\bx )-\sum_{n=1}^{\infty}\frac{(-1)^n}{n!}K_n r_{e}^{n}
\frac{\partial^n}{\partial z^n}\delta^{(3)}(\bx ) \right] \, ,
\ee 
This expansion may be compared with the expansion of $\widehat{\rho}$ given by Eq. (\ref{33}).

\subsection{Multipole structure of the world function}

The function $\Om^{(PN)}(x_A,x_B)$ is determined by 
Eqs. (\ref{25})-(\ref{25c}) where $w$ and $\bw$ are respectively replaced by
$W$ and $\bW$. The integrals involved in the r.h.s. of 
Eqs. (\ref{25})-(\ref{25c}) are given by Eqs. (\ref{28}) and (\ref{29}). 
Substituting Eq. (\ref{33}) into Eq. (\ref{28})
and using the properties of the Dirac distribution, we obtain
\be \lb{39}
\int_{0}^{1}W\left( \bx_{(0)}(\lambda )\right) d\lambda =
GM\left[ 1-\sum_{n=2}^{\infty}\frac{1}{n!}J_nr_{e}^{n}
\frac{\partial^n}{\partial z^n}\right] F(\bx ,\bx_A,\bx_B)\bigg|_{\bx =0}\, .
\ee
Similarly, substituting Eq. (\ref{38}) into Eq. (\ref{29}), we get \footnote{Note that the 
sign of Eq. (55) in Ref. \cite{linet1} is erroneous.}
\be \lb{40}
\int_{0}^{1}\bW \left(\bx_{(0)}(\lambda )\right) d\lambda =
\frac{1}{2}G\bS \times \bna \left[1-\sum_{n=1}^{\infty}\frac{1}{n!}K_n
r_{e}^{n}\frac{\partial^n}{\partial z^n}\right] F(\bx ,\bx_A,\bx_B)
\bigg|_{\bx=0} \, .
\ee

These formulae show that the multipole expansion of
$\Om^{(PN)}(x_A,x_B)$ can be thoroughly calculated by straightforward
differentiations of the kernel function $F(\bx ,\bx_A,\bx_B)$ given by 
Eq. (\ref{19}). They constitute an essential result, since they give an algorithmic 
procedure for determining the multipole expansions of the time transfer function 
and of the frequency shift in a stationary axisymmetric field (see also Ref. \cite{kop1}).

In order to obtain explicit formulae, we shall only retain the
contributions due to $M$, $J_2$ and $\bS$ in the expansion yielding
$\Om^{(PN)}_W$ and $\Om^{(PN)}_{\bW}$. Then, 
denoting the unit vector along the $z$-axis by $\bk$
and noting that $\bS=S\bk$, we get for $\Om_{W}^{(1)}(x_A,x_B)$
\bea \lb{41}
\Om_{W}^{(PN)}(x_A,x_B) & = & -\frac{GM}{c^2}
\frac{(x_{B}^{0}-x_{A}^{0})^2+\gamma R_{AB}^{2}}{R_{AB}}
\ln \left( \frac{r_A+r_B+R_{AB}}{r_A+r_B-R_{AB}}\right) \nonumber \\
&   & \mbox{} +\frac{2GM}{c^2}J_2r_{e}^{2}
\frac{(x_{B}^{0}-x_{A}^{0})^2+\gamma R_{AB}^{2}}
{\left[ (r_A+r_B)^2-R_{AB}^{2}\right]^2}(r_A+r_B)
\left( \frac{\bk \cdot \bx_A}{r_A} +
\frac{\bk \cdot \bx_B}{r_B} \right)^2 \nonumber \\
&   & \mbox{} -\frac{GM}{c^2}J_2r_{e}^{2}
\frac{(x_{B}^{0}-x_{A}^{0})^2+\gamma R_{AB}^{2}}{(r_A+r_B)^2-R_{AB}^{2}}
\left[ \frac{(\bk \times \bx_A )^2}{r_{A}^{3}}+\frac{(\bk \times \bx_B )^2}
{r_{B}^{3}} \right] +\cdots 
\eea
and for $\Om^{(PN)}_{\bW}(x_A,x_B)$
\be \lb{42}
\Om_{\bW}^{(PN)}(x_A,x_B)=\left( \gamma +1+\frac{1}{4}\alpha_1 \right)
\frac{2GS}{c^3}(x_{B}^{0}-x_{A}^{0})\frac{r_A+r_B}{r_Ar_B} 
\frac{\bk \cdot (\bx_A \times 
\bx_B)}{(r_A+r_B)^2-R_{AB}^{2}} +\cdots \,.
\ee
Finally, owing to the limit $\mid \! \alpha_1 \! \mid < 4\times 10^{-4}$ 
furnished in \cite{will}, we shall henceforth neglect all the
multipole contributions in $\Om_{\bv_r}^{(PN)}(x_A,x_B)$. Thus, we get
\be \lb{43}
\Om_{\bv_r}^{(PN)}(x_A,x_B)=\alpha_1 \frac{GM}{2c^3}(x_{B}^{0}-x_{A}^{0})
\frac{\bR_{AB}\cdot \bv_r}{R_{AB}}\ln \left( \frac{r_A+r_B+R_{AB}}
{r_A+r_B-R_{AB}}\right) +\cdots \, .
\ee

In this section and in the following one, the symbol $+\cdots$ stands for the 
contributions of higher multipole moments which are neglected. For the sake 
of brevity, when $+\cdots$ is used, we systematically omit to mention the 
symbol $O(n)$ which stands for the neglected post-Newtonian terms.  

\subsection{Time transfer function up to the order $1/c^4$}

In what follows, we put
\be \lb{n}
\bn_A=\frac{\bx_A}{r_A} , \qquad \bn_B=\frac{\bx_B}{r_B}  \, , 
\ee
and
\be \lb{Nab}
\bN_{AB} = \{N^{i}\} = \frac{\bx_B-\bx_A}{R_{AB}} \, .
\ee
Furthermore, we use systematically the identity
\be \lb{id0}
(r_A + r_B)^2 - R_{AB}^2 = 2r_A r_B (1 + \bn_A \cdot \bn_B) \, .
\ee

By substituting $R_{AB}$ for $x_B^0 - x_A^0$ into Eqs. (\ref{41})-(\ref{43}) and inserting 
the corresponding expression of $\Om^{(PN)}$ into Eq. (\ref{5n}), we get an 
expression for the time transfer function as follows  
\be \lb{44}
{\cal T}(\bx_A,\bx_B)=\frac{1}{c}R_{AB}+{\cal T}_M(\bx_A,\bx_B)+
{\cal T}_{J_2}(\bx_A,\bx_B)+{\cal T}_{\bS}(\bx_A,\bx_B)+
{\cal T}_{\bv_r}(\bx_A,\bx_B) + \cdots \, , 
\ee
where
\bea 
{\cal T}_M(\bx_A,\bx_B) & = & (\gamma +1)\frac{GM}{c^3}
\ln \left( \frac{r_A+r_B+R_{AB}}{r_A+r_B-R_{AB}}\right) , \lb{45} \\
& & \nonumber \\
{\cal T}_{J_2}(\bx_A,\bx_B) & = & -\frac{\gamma +1}{2}\frac{GM}{c^3}J_2 \,\frac{r_{e}^{2}}{r_A r_B}\frac{R_{AB}}{1 + \bn_A \cdot \bn_B} 
\left[ \left(\frac{1}{r_A} + \frac{1}{r_B}\right)\frac{(\bk \cdot \bn_A + \bk \cdot \bn_B)^2}{1 + \bn_A \cdot \bn_B}  \right. \nonumber \\
&   & \qquad \qquad \qquad \qquad \qquad \qquad \qquad \qquad \left.  
-\frac{1 - (\bk \cdot \bn_A)^2}{r_{A}}-\frac{1 - (\bk \cdot \bn_B)^2}{r_{B}} \right] , \lb{46} \\
& & \nonumber \\
{\cal T}_{\bS}(\bx_A,\bx_B) & = & -\left( \gamma +1+\frac{1}{4}\alpha_1 \right)\frac{GS}{c^4} 
\left(\frac{1}{r_A} + \frac{1}{r_B}\right)\frac{\bk \cdot (\bn_A \times \bn_B)}{1 + \bn_A \cdot \bn_B} , \lb{47} \\
& & \nonumber \\
{\cal T}_{\bv_r}(\bx_A,\bx_B) & = & - \, \alpha_1 \frac{GM}{2c^4} 
(\bN_{AB}\cdot \bv_r) \ln \left( \frac{r_A+r_B+R_{AB}}{r_A+r_B-R_{AB}}\right) \, . \lb{48}
\eea

The time transfer is thus explicitly determined up to the order $1/c^4$.
The term of order $1/c^3$ given by Eq. (\ref{45}) is the well-known Shapiro time 
delay \cite{sha}. Equations (\ref{46}) and (\ref{47}) extend results previously 
found for $\gamma = 1$ and $\alpha_1 = 0$ \cite{kli1}. However, our derivation is more 
straightforward and yields formulae which are more convenient to calculate the frequency 
shifts. As a final remark, it is worthy of note that ${\cal T}_M$
and ${\cal T}_{J_2}$ are symmetric in $(\bx_A,\bx_B)$, while ${\cal T}_{\bS}$
and ${\cal T}_{\bv_r}$ are antisymmetric in $(\bx_A,\bx_B)$.

\subsection{Directions of light rays at $x_A$ and $x_B$ 
up to the order $1/c^3$}

In order to determine the vectors tangent to the ray path at
$x_A$ and $x_B$, we use Eqs. (\ref{6}) and (\ref{5}) where ${\cal T}$ is
replaced by the expression given by Eqs. (\ref{44})-(\ref{48}). It is clear that 
$\widehat{\bl}_{A}$ and $\widehat{\bl}_{B}$ may be written as
\bea 
\widehat{\bl}_{A} &=& - \bN_{AB} + \bla_{e}(\bx_A, \bx_B) \, , \lb{bl1} \\
\widehat{\bl}_{B} &=& - \bN_{AB} + \bla_{r}(\bx_A, \bx_B) \, , \lb{bl2}
\eea
where $\bla_{e}$ and $\bla_{r}$ are perturbation terms due to ${\cal T}_{M}$, ${\cal T}_{J_{n}}$, ${\cal T}_{\bS}$, 
${\cal T}_{K_{n}}$, ... For the expansion of ${\cal T}$ given by Eqs. (\ref{44})-(\ref{48}), we find 
\be \lb{49a} 
\bla_{e}(\bx_A,\bx_B)= - \bla_M(\bx_B,\bx_A) - \bla_{J_2}(\bx_B,\bx_A)
+ \bla_{\bS}(\bx_B,\bx_A) + \bla_{\bv_r}(\bx_B,\bx_A) + \cdots \, ,
\ee
\be \lb{49b} 
\bla_{r}(\bx_A,\bx_B) = \bla_M(\bx_A,\bx_B) + \bla_{J_2}(\bx_A,\bx_B)
+ \bla_{\bS}(\bx_A,\bx_B) + \bla_{\bv_r}(\bx_A,\bx_B) + \cdots \, , 
\ee
where $\bla_M$, $\bla_{J_2}$, $\bla_{\bS}$ and $\bla_{\bv_r}$ stand for the
contributions of ${\cal T}_M$, ${\cal T}_{J_2}$, ${\cal T}_{\bS}$
and ${\cal T}_{\bv_r}$, respectively. We get from Eq. (\ref{45})
\be \lb{50}
\bla_M(\bx_A,\bx_B) = -(\gamma +1)\frac{GM}{c^2}\left(\frac{1}{r_A} + \frac{1}{r_B} \right)\frac{1}{1 + \bn_A \cdot \bn_B}
\left(\bN_{AB} - \frac{R_{AB}}{r_A + r_B}\bn_B  \right) \, .
\ee

From Eq. (\ref{46}), we get
\bea \lb{51}
& & \bla_{J_2}(\bx_A,\bx_B) = (\gamma +1)\frac{GM}{c^2} \left(\frac{1}{r_A} + \frac{1}{r_B}\right)\, J_2 \, \frac{r_{e}^{2}}{r_A r_B}
\frac{1}{\left( 1 + \bn_A \cdot \bn_B\right)^2}  \nonumber \\
& & \qquad \qquad \qquad  \; \times \left\{ \bN_{AB} 
\left[ \frac{(\bk\cdot \bn_A + \bk\cdot \bn_B)^2}{1 + \bn_A \cdot \bn_B}\left(\frac{r_A}{r_B}+ \frac{r_B}{r_A} + \frac{1}{2}
- \frac{3}{2} \bn_A \cdot \bn_B\right)\right. 
\right. \nonumber \\ 
& & \qquad \qquad \qquad \quad \left. - \frac{1}{2} \frac{r_A r_B}{r_A + r_B}\left( \frac{1-(\bk\cdot \bn_A)^2}{r_A}+
\frac{1-(\bk\cdot \bn_B)^2}{r_B}\right)\left(\frac{r_A}{r_B}+ \frac{r_B}{r_A} + 1 -\bn_A \cdot \bn_B  \right) \right] \nonumber \\
& & \qquad \qquad \qquad \quad  - \bn_B \frac{R_{AB}}{r_A + r_B}\left[ \frac{(\bk\cdot \bn_A+\bk\cdot \bn_B)^2}{1 + \bn_A \cdot \bn_B}
\left(\frac{r_A}{r_B}+ \frac{r_B}{r_A} + \frac{3}{2}
- \frac{1}{2} \bn_A \cdot \bn_B\right) \right.  \nonumber \\ 
& & \qquad \qquad \qquad \quad -\frac{1}{2}\left[ 1-3(\bk\cdot \bn_B)^2\right]  \frac{r_A ( 2 + \bn_A \cdot \bn_B )+ r_B}{r_B} 
  \nonumber \\
& & \qquad \qquad \qquad \quad \left. - \frac{1}{2}(r_A+r_B)\left( \frac{1-(\bk\cdot \bn_A)^2}{r_A} - \frac{2(\bk\cdot \bn_A)(\bk\cdot \bn_B)}{r_B}
\right) \right] \nonumber \\
& & \qquad \qquad \qquad \quad \left. + \bk \, \frac{R_{AB}}{r_B}\left[(\bk \cdot \bn_A ) + (\bk\cdot \bn_B)
\frac{ r_A (2 + \bn_A \cdot \bn_B) + r_B}{r_A+r_B}\right] \right\} \, . 
\eea

\newpage

From Eqs. (\ref{47}) and (\ref{48}), we derive the other contributions that are not neglected here :
\bea 
& & \bla_{\bS}(\bx_A,\bx_B)=\left( \gamma + 1 + \frac{1}{4}\alpha_1\right) \frac{GS}{c^3 r_B}\left(\frac{1}{r_A} + \frac{1}{r_B}\right)
\frac{1}{1+ \bn_A \cdot \bn_B} \nonumber \\
& & \qquad \qquad \qquad \qquad \qquad \qquad \quad  
\times \bigg\{ \bk \times \bn_A  - \frac{\bk \cdot (\bn_A\times \bn_B)}{1+ \bn_A \cdot \bn_B}
\left[ \bn_A + \frac{r_A (2 + \bn_A \cdot \bn_B) + r_B}{r_A+r_B}\bn_B\right] \bigg\} \, , \lb{52} \\
& & \nonumber \\
& & \bla_{\bv_r}(\bx_A,\bx_B) = \alpha_1\frac{GM}{2c^3}
\left[ \frac{\bv_r-(\bv_r\cdot \bN_{AB})
\bN_{AB}}{R_{AB}}\ln \left( \frac{r_A+r_B+R_{AB}}{r_A+r_B-R_{AB}}
\right) \right. \nonumber \\
& & \qquad \qquad \qquad \qquad \qquad \qquad \quad \left. + \frac{(\bv_r \cdot \bN_{AB})}{1 + \bn_A \cdot \bn_B}
\left(\frac{1}{r_A} + \frac{1}{r_B}\right)
\left(\bN_{AB} - \frac{R_{AB}}{r_A + r_B}\bn_B \right) \right] \, . \lb{53}
\eea

We note that the mass and the quadrupole moment yield contributions of
order $1/c^2$, while the intrinsic angular momentum and the velocity
relative to the universe rest frame yield contributions of order $1/c^3$.

\subsection{Sagnac terms in the time transfer function}

In experiments like ACES Mission, recording the time of emission $t_A$ will be more
practical than recording the time of reception $t_B$. So, it will be very
convenient to form the expression of the time transfer ${\cal T}(\bx_A , \bx_B )$ from
$\bx_A(t_A)$ to $\bx_B(t_B)$ in terms of the position of the receiver B
at the time of emission $t_A$. For any quantity $Q_B(t)$ defined along
the world line of the station B, let us put $\widetilde{Q}_B=Q(t_A)$. Thus we
may write $\widetilde{\bx}_B(t_A)$, $\widetilde{r}_B(t_A)$, $\widetilde{\bv}_B(t_A)$, 
$\widetilde{v}_B = \mid \! \widetilde{\bv}_B \! \mid$, etc.

Now, let us introduce the instantaneous coordinate distance 
$\bD_{AB}=\widetilde{\bx}_B-\bx_A$ and its norm $D_{AB}$. Since we want 
to know $t_B-t_A$ up to the order $1/c^4$, we can use the Taylor expansion of
$\bR_{AB}$
$$
\bR_{AB}=\bD_{AB}+(t_B-t_A)\widetilde{\bv}_B+\frac{1}{2}(t_B-t_A)^2 \,
\widetilde{\ba}_B+\frac{1}{6}(t_B-t_A)^3 \, \widetilde{\bb}_B+\cdots \, ,
$$
where $\ba_B$ is the acceleration of B and $\bb_B=d\ba_B/dt$. Using 
iteratively this expansion together with Eq. (\ref{44}), we get
\bea \lb{55}
& & {\cal T}(\bx_A,\bx_B)={\cal T}(\bx_A,\widetilde{\bx}_B)+ 
\frac{1}{c^2}\bD_{AB} \cdot \widetilde{\bv}_B 
+\frac{1}{2c^3}D_{AB} \left[ \frac{(\bD_{AB}\cdot \widetilde{\bv}_B)^2}
{D_{AB}^{2}}+\widetilde{v}_{B}^{2}+\bD_{AB}\cdot \widetilde{\ba}_B\right] \nonumber \\
& & \qquad \qquad \qquad +\frac{1}{c^4}\left[ \left( \bD_{AB}\cdot 
\widetilde{\bv}_B\right) \left( \widetilde{v}_{B}^{2}
+\bD_{AB}\cdot \widetilde{\ba}_B \right) +\frac{1}{2}D_{AB}^{2} \left(
\widetilde{\bv}_B\cdot \widetilde{\ba}_B+\frac{1}{3}\bD_{AB}\cdot \widetilde{\bb}_B \right)
\right] \nonumber \\
& & \qquad \qquad \qquad +\frac{1}{c}\frac{\bD_{AB}}{D_{AB}}\cdot 
\widetilde{\bv}_B \left[ {\cal T}_M(\bx_A,\widetilde{\bx}_B)+{\cal T}_{J_2}(\bx_A,\widetilde{\bx}_B) \right] 
\nonumber \\
& & \qquad \qquad \qquad - \frac{1}{c^2}D_{AB}\widetilde{\bv}_B \cdot \left[ \bla_M(\bx_A, \widetilde{\bx}_B)
+\bla_{J_2}(\bx_A, \widetilde{\bx}_B)\right] + \cdots \, , 
\eea
where ${\cal T}(\bx_A,\widetilde{\bx}_B)$ is obtained by substituting 
$\widetilde{\bx}_B$, $\widetilde{r}_B$ and $\bD_{AB}$ respectively for $\bx_B$, $r_B$ 
and $\bR_{AB}$ into the time transfer function defined by 
Eqs. (\ref{44})-(\ref{48}). This expression extends the previous formula \cite{bla1} to the 
next order $1/c^4$. The second, the third
and the fourth terms in Eq. (\ref{55}) represent pure Sagnac terms
of order $1/c^2$, $1/c^3$ and $1/c^4$, respectively. The fifth and the sixth terms are 
contributions of the gravitational field mixed with the coordinate velocity of the 
receiving station. Since these last two terms are of order $1/c^4$, they might be 
calculated for the arguments $(\bx_A,\bx_B)$.

\section{Frequency shift in the field of a rotating axisymmetric body}

\subsection{General formulae up to the fourth order}

It is possible to derive the ratio $q_A/q_B$ up to the order $1/c^4$
from our results in Sec. IV since $\widehat{\bl}_A$ and $\widehat{\bl}_B$ are given up to
the order $1/c^3$ by Eqs. (\ref{bl1})-(\ref{49b}). Denoting by
$\widehat{\bl}^{(n)}/c^n$ the $O(n)$ terms in $\widehat{\bl}$, $q_A/q_B$ may be expanded as
\bea \lb{59}
\frac{q_A}{q_B} &=& 1-\frac{1}{c}\frac{\bN_{AB}\cdot (\bv_A-\bv_B)}
{1 \displaystyle - \bN_{AB}\cdot \frac{\bv_B}{c}} 
+\frac{1}{c^3}\left[ \widehat{\bl}_{A}^{(2)}\cdot \bv_A-\widehat{\bl}_{B}^{(2)}\cdot \bv_B 
\right] +\frac{1}{c^4}\left[ \widehat{\bl}_{A}^{(3)}\cdot \bv_A-\widehat{\bl}_{B}^{(3)}
\cdot \bv_B \right] \nonumber \\
& &\mbox{}  +\frac{1}{c^4}\bN_{AB}\cdot \left[ \left( \widehat{\bl}_{B}^{(2)}\cdot \bv_B\right)
(\bv_A-2\bv_B)+\left( \widehat{\bl}_{A}^{(2)}\cdot \bv_A 
\right) \bv_B \right]+ O(5)  \, . 
\eea 

In order to be consistent with this expansion, we have to perform the
calculation of $u_{A}^{0}/u_{B}^{0}$ at the same level of approximation.
For a clock delivering a proper time $\tau$, $1/u^0$ is the ratio of the
proper time $d\tau$ to the coordinate time $dt$. To reach the suitable
accuracy, it is therefore necessary to take into account the terms
of order $1/c^4$ in $g_{00}$. For the sake of simplicity, we shall henceforth 
confine ourselves to the fully conservative metric theories of gravity 
without preferred location effects, in which all the PPN parameters vanish 
except $\beta$ and $\gamma$. Since the gravitational field is assumed to be
stationary, the chosen coordinate system is then a standard 
post-Newtonian gauge and the metric reduces to its usual form 
\be \lb{M15}
G_{00}=1-\frac{2}{c^2}W+\frac{2\beta}{c^4}W^2+O(6) , \quad
\{G_{0i}\}=\frac{2(\gamma +1)}{c^3}\bW +O(5) , \quad
G_{ij}=-\left( 1+\frac{2\gamma}{c^2}W \right) \delta_{ij}+O(4) \, ,
\ee
where $W$ given by Eq. (\ref{M10}) reduces to
\be \lb{M18}
W(\bx )=U(\bx )+\frac{G}{c^2}\int \frac{\rho^*(\bxp )}{\mid \! \bx -\bxp \! \mid}
\left[ \left( \gamma +\frac{1}{2}\right) v^2+(1-2\beta )U+\Pi
+3\gamma \frac{p}{\rho^*}\right] d^3 \bxp \, ,
\ee
and $\bW$ is given by Eq. (\ref{34b}). As a consequence, for a clock moving with 
the coordinate velocity $\bv$, the quantity $1/u^0$ is given by the formula
\be \lb{61}
\frac{1}{u^0}\equiv \frac{d\tau}{dt}=1-\frac{1}{c^2}\left( W+
\frac{1}{2}v^2\right) 
+\frac{1}{c^4}\left[ \left( \beta -\frac{1}{2} \right) W^2
-\left( \gamma +\frac{1}{2}\right) Wv^2-\frac{1}{8}v^4
+2(\gamma +1)\bW\cdot \bv \right] +O(6) \, ,
\ee
from which it is easily deduced that
\bea 
\frac{u_{A}^{0}}{u_{B}^{0}} = 1 & + & \frac{1}{c^2}\left( W_A-W_B+
\frac{1}{2}v_{A}^{2}-\frac{1}{2}v_{B}^{2}\right) \nonumber \\ 
& + & \frac{1}{c^4}\bigg\{ (\gamma +1)(W_Av_{A}^{2}-W_Bv_{B}^{2})
+\frac{1}{2}(W_A-W_B)\left[ W_A-W_B  + 2(1-\beta )
(W_A+W_B) + v_{A}^{2} - v_{B}^{2} \right]  \nonumber \\
&   & \mbox{} - 2(\gamma + 1 )(\bW_A\cdot \bv_A -\bW_B\cdot \bv_B)
+ \frac{3}{8}v_{A}^{4}-\frac{1}{4}v_{A}^{2}v_{B}^2
-\frac{1}{8}v_{B}^{4}\bigg\}
+O(6) \, . \lb{65}
\eea
  
It follows from Eq. (\ref{59}) and Eq. (\ref{65}) that the frequency shift
$\delta \nu /\nu$ is given by
\be \lb{66}
\frac{\delta \nu}{\nu}\equiv
\frac{\nu_A}{\nu_B}-1=\left( \frac{\delta \nu}{\nu}\right)_c +
\left( \frac{\delta \nu}{\nu}\right)_g \, ,
\ee
where $(\delta \nu /\nu )_c$ is the special-relativistic Doppler effect
\bea 
& & \left( \frac{\delta \nu}{\nu}\right)_c = 
-\frac{1}{c}\bN_{AB}\cdot (\bv_A-\bv_B) 
+\frac{1}{c^2}\left[ \frac{1}{2}v_{A}^{2}-\frac{1}{2}v_{B}^{2}-
\left( \bN_{AB}\cdot (\bv_A-\bv_B)\right) \left( \bN_{AB}\cdot \bv_B \right)
\right] \nonumber \\
& & \qquad \qquad -\frac{1}{c^3}\left[ \left( \bN_{AB}\cdot (\bv_A-\bv_B)\right)
\left( \frac{1}{2}v_{A}^{2}-\frac{1}{2}v_{B}^{2}+
\left( \bN_{AB}\cdot \bv_B\right)^2 \right) \right] \nonumber \\
& & \qquad \qquad +\frac{1}{c^4}\left[ \frac{3}{8}v_{A}^{4}-\frac{1}{4}v_{A}^{2}v_{B}^{2}
-\frac{1}{8}v_{B}^{4} \right. \nonumber \\
& & \qquad \qquad \qquad \left. -\left( \bN_{AB}\cdot (\bv_A-\bv_B)\right) \left( \bN_{AB}\cdot 
\bv_B \right) \left( \frac{1}{2}v_{A}^{2}-\frac{1}{2}v_{B}^{2}
+\left( \bN_{AB}\cdot \bv_B \right)^2 \right) \right] +O(5) \lb{67}
\eea  
and $(\delta \nu )/\nu )_g$ contains all the contribution of the
gravitational field, eventually mixed with kinetic terms
\bea \lb{68}
& & \left( \frac{\delta \nu}{\nu}\right)_g=\frac{1}{c^2}(W_A-W_B) 
-\frac{1}{c^3}\left[ (W_A-W_B)\left( \bN_{AB}\cdot (\bv_A-\bv_B)\right) -
\widehat{\bl}_{A}^{(2)}\cdot \bv_A+\widehat{\bl}_{B}^{(2)}\cdot \bv_B \right] \nonumber \\
& & \qquad \qquad +\frac{1}{c^4}\left\{ (\gamma +1)(W_Av_{A}^{2}-W_Bv_{B}^{2})
+\frac{1}{2}(W_A-W_B)\left[ W_A-W_B+2(1-\beta )(W_A+W_B)+v_{A}^{2}-v_{B}^{2} 
\right. \right. \nonumber \\
& & \qquad \qquad \left. \left. -2\left( \bN_{AB}\cdot (\bv_A-\bv_B) \right) 
\left( \bN_{AB}\cdot \bv_B \right) \right]
+ \bN_{AB}\cdot \left[ \left( \widehat{\bl}_{B}^{(2)}\cdot \bv_B \right) 
(\bv_A-2\bv_B)+\left( \widehat{\bl}_{A}^{(2)}\cdot \bv_A \right) \bv_B \right] 
\right.  \nonumber \\ 
& & \qquad \qquad \left. +\left( \widehat{\bl}_{A}^{(3)}-2(\gamma +1)
\bW_A \right)\cdot \bv_A
-\left( \widehat{\bl}_{B}^{(3)}-2(\gamma +1)\bW_B \right) \cdot \bv_B \right\}
+O(5)  \, .
\eea

It must be emphasized that the formulae Eqs. (\ref{61}) and (\ref{65}) are valid within the PPN
framework without adding special assumption, provided that $\beta$ and
$\gamma$ are the only non-vanishing post-Newtonian parameters. On the
other hand, Eq. (\ref{68}) is valid only for stationary gravitational fields.
In the case of an axisymmetric rotating body, we shall obtain an
approximate expression of the frequency shift by inserting the following
developments in Eq. (\ref{68}), yielded by Eqs. (\ref{49a})-(\ref{53}):   
\bea
& & \widehat{\bl}_{A}^{(2)}/c^2= - \bla_M(\bx_B,\bx_A)- \bla_{J_2}(\bx_B,\bx_A)+ \cdots 
\, , \quad \,
\widehat{\bl}^{(3)}_{A}/c^3= \bla_{\bS}(\bx_B,\bx_A)+ \cdots \, , \nonumber \\
& & \widehat{\bl}_{B}^{(2)}/c^2 = \bla_M(\bx_A,\bx_B) + \bla_{J_2}(\bx_A,\bx_B) 
+\cdots \, , \qquad
\widehat{\bl}^{(3)}_{B}/c^3 = \bla _{\bS}(\bx_A,\bx_B)+ \cdots \, , \nonumber 
\eea
the function $\bla_{\bS}$ being now given by Eq. (\ref{52}) written with
$\alpha_1=0$. Let us recall that the symbol $+ \cdots$ stands for the contributions 
of the higher multipole moments which are neglected.

\subsection{Application in the vicinity of the Earth}

In order to perform numerical estimates of the frequency shifts in the 
vicinity of the Earth, we suppose now that A is on board the International
Space Station (ISS) orbiting at the altitude $H=400$ km and that B is a
terrestrial station. It will be the case for the ACES mission. We use 
$r_B = 6.37 \times 10^6$ m and $r_A - r_B =400$ km. For 
the velocity of ISS, we take $v_A = 7.7 \times 10^3$ m/s and for the 
terrestrial station, we have $v_B \leq 465$ m/s. The other useful parameters 
concerning the Earth are: $G M = 3.986 \times 10^{14}$ m$^3$/s$^2$, 
$r_e = 6.378 \times 10^6$ m, $J_2 = 1.083 \times 10^{-3}$; for 
$n\geq 3$, the multipole moments $J_n$ are in the order of $10^{-6}$. With 
these values, we get $W_B/c^2 \approx G M /c^2 r_B = 6.95 \times 10^{-10}$ and 
$W_A /c^2 \approx  G M /c^2 r_A = 6.54 \times 10^{-10}$. From these data, it is 
easy to deduce the following upper bounds:
$\mid \! \bN_{AB} \cdot \bv_A /c \! \mid \leq  2.6 \times 10^{-5}$ 
for the satellite, 
$\mid \! \bN_{AB}\cdot \bv_B /c \! \mid \leq 1.6 \times 10^{-6}$ 
for the ground station and 
$\mid \! \bN_{AB}\cdot (\bv_A-\bv_B)/c \! \mid \leq 2.76 \times 10^{-5}$
for the first-order Doppler term.

Our purpose is to obtain correct estimates of the effects
in Eq. (\ref{68}) with are greater than or equal to $10^{-18}$ for an axisymmetric
model of the Earth. At this level of approximation, it is not sufficient
to take into account the $J_2$-terms in $(W_A-W_B)/c^2$. First, the
higher-multipole moments $J_3$, $J_4$, $\ldots$ yield contribution of order
$10^{-15}$ in $W_A/c^2$. Second, owing to the irregularities in the
distribution of masses, the expansion of the geopotential in a series of
spherical harmonics is probably not convergent at the surface of the
Earth. For these reasons, we do not expand $(W_A-W_B)/c^2$ in Eq. (\ref{68}).

However, for the higher-order terms in Eq. (\ref{68}), we can apply the 
explicit formulae obtained in the previous section. Indeed, since the
difference between the geoid and the reference ellipsoid is less than
$100$ m, $W_B/c^2$ may be written as \cite{wolf}
\[
\frac{1}{c^2}W_B=\frac{GM}{c^2r_B}+\frac{GMr_{e}^{2}J_2}{2c^2 r_{B}^{3}}
(1-3\cos^2\theta )+\frac{1}{c^2}\triangle W_B \, ,
\]
where the residual term $\triangle W_B/c^2$ is such that 
$\mid \! \triangle W_B/c^2 \! \mid \leq 10^{-14}$.
At a level of experimental uncertainty about $10^{-18}$, this inequality 
allows to retain only the contributions due to $M$,
$J_2$ and $\bS$ in the terms of orders $1/c^3$ and $1/c^4$. As a
consequence, the formula (\ref{68}) reduces to  
\bea \lb{0665}
& & \left(\frac{\delta \nu}{\nu} \right)_g=\frac{1}{c^2}(W_A-W_B)+
\frac{1}{c^3}\left( \frac{\delta \nu}{\nu} \right)_{M}^{(3)} + 
\frac{1}{c^3}\left( \frac{\delta \nu}{\nu} \right)_{J_2}^{(3)} 
+\cdots \nonumber \\ 
& & \qquad \qquad +\frac{1}{c^4}\left( \frac{\delta \nu}{\nu} \right)_{M}^{(4)}+ 
\frac{1}{c^4}\left( \frac{\delta \nu}{\nu} \right)_{\bS}^{(4)}+\cdots \, , 
\eea
where the different terms involved in the r.h.s. are separately explicited and
discussed in what follows.

By using Eq. (\ref{id0}), it is easy to see that $(\delta \nu /\nu )_{M}^{(3)}$ is given by
\bea \lb{06652}
& & \left( \frac{\delta \nu}{\nu} \right)_{M}^{(3)}= - \frac{GM(r_A+r_B)}{r_Ar_B}
\left[ \left( \frac{\gamma +1}{1+\bn_A\cdot \bn_B}-\frac{r_A-r_B}{r_A+r_B}
\right) [\bN_{AB}\cdot (\bv_A-\bv_B)] \right. \nonumber \\
& & \qquad \qquad \; \left. +(\gamma +1)\frac{R_{AB}}{r_A+r_B}\frac{\bn_A\cdot \bv_A+\bn_B
\cdot \bv_B}{1+\bn_A\cdot \bn_B}\right] \, .
\eea
The contribution of this term is bounded by $5\times 10^{-14}$
for $\gamma =1$, in accordance with a previous analysis \cite{bla1}.

\subsection{Influence of the quadrupole moment at the order $1/c^3$}

It follows from Eqs. (\ref{51}) and (\ref{68}) that the term 
$\left(\delta \nu /\nu \right)_{J_{2}}^{(3)}$ in Eq. (\ref{0665}) is given by
\bea \lb{0741}
\left(\frac{\delta \nu}{\nu} \right)_{J_2}^{(3)} & = & \frac{G M}{2r_e} J_{2} 
\left( \bN_{AB} \cdot (\bv_A - \bv_B ) \right) \left[
\left( \frac{r_e}{r_A} \right)^3 \left[ 3 (\bk \cdot \bn_A)^2 - 1 \right] - 
\left(\frac{r_e}{r_B} \right)^3 
\left[ 3 (\bk \cdot \bn_B)^2 - 1 \right] \right] \nonumber \\ 
&  & \mbox{} + \, (\gamma +1) 
GM \left(\frac{1}{r_A} + \frac{1}{r_B}\right) J_2 \, \frac{r_{e}^{2}}{r_A r_B}\, \frac{1}{(1 + \bn_A \cdot \bn_B)^2}  \nonumber \\   
&  & \mbox{} \times \bigg\{ [\bN_{AB}\cdot (\bv_A - \bv_B)]
\left[ \frac{(\bk \cdot \bn_A + \bk \cdot \bn_B )^2}{1 + \bn_A . \bn_B} \,\left( \frac{r_A}{r_B} + \frac{r_B}{r_A} 
+ \frac{1}{2} - \frac{3}{2} \bn_A \cdot \bn_B \right) \right.  \nonumber \\
& & \qquad  \left. - \, \frac{1}{2}\left( 1 - \frac{r_A (\bk \cdot \bn_B)^2 + 
r_B (\bk \cdot \bn_A)^2}{r_A +r_B} \right) 
\left(\frac{r_A}{r_B} + \frac{r_B}{r_A} + 1 - \bn_A \cdot \bn_B \right) \right]   \nonumber \\
& & \qquad  + \, \frac{R_{AB}}{r_A + r_B} (\bn_A \cdot \bv_A + \bn_B \cdot \bv_B ) 
\frac{(\bk \cdot \bn_A + \bk \cdot \bn_B )^2 }{1 + \bn_A \cdot \bn_B}\,   
\left( \frac{r_A}{r_B} + \frac{r_B}{r_A} 
+ \frac{3}{2} - \frac{1}{2} \bn_A \cdot \bn_B \right)  \nonumber \\ 
& & \qquad  - \, \frac{1}{2}\frac{R_{AB}}{r_A}( \bn_A \cdot \bv_A ) \left[1 - 3(\bk \cdot \bn_A)^2 \right] 
\frac{r_A +  r_B (2 + \bn_A \cdot \bn_B)}{r_A + r_B} \nonumber \\
& & \qquad  - \, \frac{1}{2}\frac{R_{AB}}{r_B} ( \bn_B \cdot \bv_B ) \left[1 - 3(\bk \cdot \bn_B)^2 \right]  
\frac{r_A (2 + \bn_A \cdot \bn_B ) + r_B}{r_A + r_B} \nonumber \\
& & \qquad  + \, R_{AB} \left[  \left( \frac{\bn_A \cdot \bv_A}{r_A} + \frac{\bn_B \cdot \bv_B}{r_B} \right)
(\bk \cdot \bn_A)(\bk \cdot \bn_B) \right. \nonumber \\
& &  \qquad \qquad \quad \; \; \left. - \, \frac{1}{2}(\bn_A \cdot \bv_A ) \frac{1-(\bk \cdot \bn_B)^2}{r_B} 
 - \, \frac{1}{2}(\bn_B \cdot \bv_B ) \frac{1 - (\bk \cdot \bn_A)^2}{r_A} \right] \nonumber \\
& &  \qquad  - \, \frac{R_{AB}}{r_A }( \bk \cdot \bv_A ) \left[ \bk \cdot \bn_A 
\frac{r_A + r_B(2 + \bn_A \cdot \bn_B )}{r_A + r_B}  + \bk \cdot \bn_B \right]   \nonumber \\
& &  \qquad  - \, \frac{R_{AB}}{r_B}( \bk \cdot \bv_B ) \left[ \bk \cdot \bn_A  
+ \bk \cdot \bn_B  \frac{r_A (2 + \bn_A \cdot \bn_B ) + r_B}{r_A + r_B} \right] \bigg\} \, . 
\eea 

One has $\mid \bv_A /c \mid = 2.6 \times 10^{-5}$, 
$\mid \bv_B /c \mid \leq 1.6 \times 10^{-6} $ and $K_{AB}=3.77 \times 10^{-3}$. A crude estimate 
can be obtained by neglecting in (\ref{0741}) the terms involving the scalar products $\bn_B \cdot \bv_B$
and $\bk \cdot \bv_B$. Since the orbit of ISS is almost circular, the scalar product 
$\bn_A \cdot \bv_A$ can also be neglected. On these assumptions, we find for $\gamma =1$
\be \lb{07414}
\bigg| \frac{1}{c^3} \left( \frac{\delta \nu}{\nu}\right)_{J_2}^{(3)} \bigg| \leq  1.3 \times 10^{-16} .
\ee 

As a consequence, it will probably be necessary to take into account the $O(3)$ contributions of $J_{2}$ 
in the ACES mission. This conclusion is to be compared 
with the order of magnitude given in \cite{bla1} without a detailed calculation. Of course, a better 
estimate might be found if the inclination $i = 51.6 \deg$ of the 
orbit with respect to the terrestrial 
equatorial plane and the latitude $\pi /2 - \theta_B$ of the 
ground station were taken into account.

\subsection{Frequency shifts of order $1/c^4$}

The term $(\delta \nu /\nu)_{M}^{(4)}$ in Eq. (\ref{0665}) is given by
\bea 
& & \left(\frac{\delta \nu}{\nu} \right)_{M}^{(4)} =  
(\gamma + 1)\left(\frac{GM}{r_A}v_A^2 - \frac{GM}{r_B}v_B^2 \right) 
-\frac{GM (r_A-r_B)}{2 \, r_A r_B}
(v_A^2 - v_B^2) \nonumber \\
& & \qquad \qquad + \, \frac{1}{2} \left(\frac{GM}{r_A r_B}\right)^2 
\left[(r_A-r_B )^2+2(\beta -1)(r_A^2 -r_B^2) \right]  \nonumber \\
& & \qquad \qquad \; - \, \frac{GM(r_A+r_B)}{r_Ar_B}\left[ \left( \frac{2(\gamma +1 )}
{1+\bn_A\cdot \bn_B}-\frac{r_A-r_B}{r_A+r_B}\right) [ \bN_{AB}
\cdot (\bv_A-\bv_B)] \left( \bN_{AB}\cdot \bv_B \right) 
\right. \nonumber \\
& & \qquad \qquad \; \left. + \,\frac{\gamma + 1}{1+\bn_A\cdot \bn_B}
\frac{R_{AB}}{r_A+r_B}
\{(\bn_A\cdot \bv_A)\left( \bN_{AB}\cdot \bv_B\right)
- [\bN_{AB}\cdot (\bv_A - 2\bv_B)](\bn_B\cdot \bv_B)\} \right] \lb{06652a} \, .
\eea
The dominant term $(\gamma + 1) GM v_A^2/r_A$ in Eq. (\ref{06652a}) induces a correction to the frequency
shift which amounts to $10^{-18}$. So, it will certainly be necessary to take 
this correction into account in experiments performed in the foreseeable future.

The terms $(\delta \nu /\nu)_{\bS}^{(4)}$ is the contribution of the 
intrinsic angular momentum to the frequency shift. 
Substituting Eqs. (\ref{36}) and (\ref{52}) into Eq. (\ref{68}), it may be seen that
\be \lb{075}
\left( \frac{\delta \nu}{\nu} \right)_{\bS}^{(4)}  = 
\left({\cal F}_{\bS}  \right)_A- \left({\cal F}_{\bS} \right)_B \, ,
\ee 
where
\bea 
\left({\cal F}_{\bS} \right)_A & = &  (\gamma + 1)\, \frac{GS }{r_{A}^{2}} 
\left(1 + \frac{r_A}{r_B} \right)\bv_{A} \cdot 
\left\{ \frac{\bk \times \bn_B}{1 + \bn_A \cdot \bn_B}  
- \frac{r_B}{r_A + r_B} \, \bk \times \bn_A  \right.  \nonumber \\
          &   & \mbox{} + \left. \frac{\bk\cdot (\bn_A \times \bn_B)}{(1 + \bn_A \cdot \bn_B)^2 }
\left[  \frac{r_A +  r_B(2 + \bn_A \cdot \bn_B)}{r_A + r_B} \bn_A  + \bn_B \right] \right\}  \, , \lb{078}
\eea
\bea 
\left({\cal F}_{\bS } \right)_B & = &  (\gamma + 1) \, \frac{GS}{r_{B}^{2}} 
\left(1+ \frac{r_B}{r_A} \right) \bv_{B} \cdot 
\left\{ \frac{\bk \times \bn_A}{1 +  \bn_A \cdot \bn_B } 
- \frac{r_A}{r_A + r_B}\, \bk \times \bn_B  \right. \nonumber \\
          &   & \mbox{} - \left. \frac{\bk \cdot (\bn_A \times \bn_B)}{(1 + \bn_A \cdot \bn_B)^2 }
\left[ \bn_A +  \frac{r_A (2 + \bn_A \cdot \bn_B) + r_B}{r_A + r_B} \bn_B \right] \right\}  \, . \lb{079}
\eea

In order to make easier the discussion, it is useful to introduce the angle $\psi$ 
between $\bx_A$ and $\bx_B$ and the angle $i_p$ between the plane of the 
photon path and the equatorial plane. These angles are defined by 
\[
\cos \psi = \bn_A \cdot \bn_B \, , \quad 0\leq \psi < \pi \, , \quad
\bk\cdot (\bn_A \times \bn_B) = \sin \psi \cos i_p \, , \quad
0 \leq i_p < \pi \, .
\]
With these definitions, it is easily seen that
\[
\frac{\bk\cdot (\bn_A \times \bn_B)}{1 + \bn_A \cdot \bn_B } = 
\cos i_p \tan \frac{\psi}{2} \, .
\]
Let us apply our formulas to ISS. Due to the 
inequality $v_B / v_A \leq 6 \times 10^{-2}$, the term 
$\left({\cal F}_{\bS} \right)_B$ in Eq. (\ref{075}) may be  neglected. From Eq. (\ref{078}), 
it is easily deduced that
\[
\mid \left({\cal F}_{\bS} \right)_A \mid \leq \, (\gamma + 1)\, 
\frac{GS}{r_{A}^{2}} \,  
\left(1+ \frac{r_{A}}{ r_{B}} \right) \,  \frac{2 + 3 
\mid \! \tan  \psi /2 \! \mid }{\mid \! 1 + \cos \psi \! \mid } 
\, v_{A} \, .
\]
Assuming $0 \leq \psi \leq \pi /2$, we have 
$(2+3\mid \! \tan \psi /2 \! \mid )/ \mid \! 1+\cos \psi \! \mid \leq 5$. 
Inserting this inequality in the previous one and taking for the Earth 
$G S / c^3 r_{A}^{2} = 3.15 \times 10^{-16}$, we find 
\be \lb{084}
\bigg| \frac{1}{c^4}\left( \frac{\delta \nu}{\nu}\right)_{\bS}^{(4)} \bigg| 
\leq \, (\gamma +1)\times 10^{-19} \, . 
\ee

Thus, we get an upper bound which 
is slightly greater than the one estimated by retaining only the term 
$h_{0i}v^i/c$ in Eq. (\ref{65}). However, our formula confirms that the intrinsic 
angular momentum of the Earth will not affect the ACES experiment.

\section{Conclusion}

It is clear that the world function $\Om (x_A, x_B)$ constitutes a powerful tool 
for determining the time delay and the frequency shift of electromagnetic 
signals in a weak gravitational field. The analytical derivations 
given here are obtained within the Nordtvedt-Will PPN formalism. We have found the general
expression of $\Om (x_A,x_B)$ up to the order $1/c^3$. This result yields 
the expression of the time transfer functions ${\mathcal T}_e(t_{A}, \bx_{A}, \bx_{B})$ and 
${\mathcal T}_r(t_{B}, \bx_{A}, \bx_{B})$
up to the order $1/c^4$. We point out that 
$\gamma$ and $\alpha_1$ are the only post-Newtonian parameters involved
in the expressions of the world-function and of the time transfer functions 
within the limit of the considered approximation.

We have treated in detail the case of an isolated, axisymmetric rotating body,
assuming that the gravitational field is stationary and that the body
is moving with a constant velocity $\bv_r$ relative to the universe rest 
frame. We have given a systematic procedure for calculating the terms due to
the multipole moments in the world function $\Om (x_A,x_B)$ and in the single time 
transfer function ${\cal T}(\bx_A,\bx_B)$. These terms are 
obtained by straightforward differentiations of a kernel function. 
We have explicitly derived the contributions due to the mass $M$, to the
quadrupole moment $J_2$ and to the intrinsic angular momentum $\bS$ of
the rotating body.

Assuming for the sake of simplicity that only $\beta$ and $\gamma$ 
are different from zero, we have determined the general expression of
the frequency shift up to the order $1/c^4$. We have derived an explicit formula 
for the contributions of $J_2$ at the order $1/c^3$. Our method would give as 
well the quadrupole contribution at the order $1/c^4$ in case of necessity. 
Furthermore, we have obtained a thorough expression for the contribution of the mass monopole 
at the fourth order, as well as the contribution of the intrinsic angular momentum $\bS$, 
which is also of order $1/c^4$. It must be pointed out that our calculations give also the 
vectors tangent to the light ray at the emission and reception points. So, our results could 
be used for determining the contributions of $J_2$ and $\bS$ to the deflection of light.  

On the assumption that the gravitational field is stationary, our formulae yield all 
the gravitational corrections to the frequency shifts up to $10^{-18}$ in the 
vicinity of the Earth. Numerically, the influence of the Earth quadrupole moment 
at the order $1/c^3$ is in the region of $10^{-16}$ for a 
clock installed on board ISS and compared with a ground-clock. As a consequence, 
this effect will probably be observable during the ACES mission. We also note that 
the leading term in the fourth-order frequency shift due 
to the mass monopole is equal to $10^{-18}$ for a clock installed on board ISS 
and compared with a ground-clock. As a consequence, this effect could be 
observable in the foreseeable future with atomic clocks using optical transitions.


\end{document}